%% file: styler.tex
\documentclass[a4paper]{article}

\usepackage{INTERSPEECH2021}

\usepackage{algorithm}
\usepackage{algpseudocode}
\usepackage{threeparttable}
\usepackage{multirow}

\newcommand{\ours}{STYLER}
\newcommand{\linearmelscaling}{Mel Calibrator}
\newcommand{\doubledecoding}{Residual Decoding}

\title{\ours: Style Factor Modeling with Rapidity and Robustness via Speech Decomposition for Expressive and Controllable Neural Text to Speech}
\name{Keon Lee, Kyumin Park, Daeyoung Kim}
\address{
  School of Computing, KAIST, Daejeon, Republic of Korea}
\email{\{keon.lee, pkm9403, kimd\}@kaist.ac.kr}

\begin{document}

\maketitle

\input{00_abstract}
\input{01_introduction}

\input{02_method_res}
\input{03_experiment}
\input{04_conclusion}
\input{05_ack}

\bibliographystyle{IEEEtran}
\bibliography{mybib}

\end{document}

%% file: 00_abstract.tex
\begin{abstract}
Previous works on neural text-to-speech (TTS) have been addressed on limited speed in training and inference time, robustness for difficult synthesis conditions, expressiveness, and controllability. Although several approaches resolve some limitations, there has been no attempt to solve all weaknesses at once. In this paper, we propose {\ours}, an expressive and controllable TTS framework with high-speed and robust synthesis. Our novel audio-text aligning method called {\linearmelscaling} and excluding autoregressive decoding enable rapid training and inference and robust synthesis on unseen data. Also, disentangled style factor modeling under supervision enlarges the controllability in synthesizing process leading to expressive TTS. On top of it, a novel noise modeling pipeline using domain adversarial training and {\doubledecoding} empowers noise-robust style transfer, decomposing the noise without any additional label. Various experiments demonstrate that {\ours} is more effective in speed and robustness than expressive TTS with autoregressive decoding and more expressive and controllable than reading style non-autoregressive TTS. Synthesis samples and experiment results are provided via our demo page\footnote{https://keonlee9420.github.io/STYLER-Demo/}, and code is available publicly\footnote{https://github.com/keonlee9420/STYLER}.
\end{abstract}

%% file: 01_introduction.tex
\section{Introduction}

Despite remarkable improvement in neural text-to-speech (TTS) towards human-level reading performance \cite{wang2017tacotron, shen2018natural, prenger2019waveglow, kong2020hifi}, criticism has been raised for the lack of expressiveness and controllability. Since the style of synthesized speech is determined from the average style of speech in the training dataset \cite{hodari2019using, stanton2018predicting}, the TTS models have been limited to represent expressive voice. In order to resolve such a problem, several approaches \cite{wang18gst, zhang19vaetaco, hsu2018hierarchical, lee2019robust, sun2020fully} has been explored so that they can synthesize and control speech with various style. 

However, the expressive TTS models have weaknesses in speed and robustness due to autoregressive architecture \cite{ren2019fastspeech}. Since each frame needs to iterate of all previous time steps to be predicted, decoding consumes significant training and inference overhead, and collapse in one step may fall into the failure of total synthesis \cite{bengio2015scheduled, ranzato2015sequence}. Also, their unsupervised style modeling reports difficult training and has a fundamental limitation on disentangling features \cite{locatello2019challenging}.

Meanwhile, speech synthesis frameworks without autoregressive decoding have been proposed recently \cite{ren2019fastspeech, peng2020non, ren2020fastspeech2}. Utilizing Transformer \cite{vaswani2017attention}, some non-autoregressive TTS models use self-attention block for parallel decoding \cite{ren2019fastspeech, ren2020fastspeech2}. The substitution of autoregressive decoder into duration prediction with supervision shows faster speed and enhances stability. Predicting value only from the single text input, however, these models still have limitations on low expressiveness and weak controllability where the influence of audio decreases.

Even with several advances in previous works, none of them covered all criticized problems at once: speed, robustness, expressivity, and controllability. To achieve this, the following conditions should be satisfied. First, autoregressive architecture needs to be avoided, which induces speed and robustness weaknesses. Second, style factor modeling from source audio should be introduced to enable more expressive and controllable synthesis, enjoying the opportunity of input variants in addition to a single text.

In this paper, we propose \textit{\ours}, a fast and robust style modeling TTS framework for expressive and controllable speech synthesis. Upon non-autoregressive decoding, we construct style factor modeling to express and control each style factor. In addition to text input, {\ours} takes and splits source audio into five components: duration, pitch, energy, content, and noise. Inspired by speech decomposition via information bottleneck \cite{qian2019autovc, qian2020speechsplit} in recent voice conversion task, our model separately encodes each style factor and decodes them into speech. 
Furthermore, our novel noise modeling decomposes noise from reference speech with no labels while successfully encoding the speech style from noisy input. To the best of our knowledge, {\ours} is the first approach of TTS framework equipped with rapidity, robustness, expressivity, and controllability at the same time with high naturalness.

The contributions of {\ours} are as follows.
\begin{itemize}
    \item {\ours} reveals faster and more robust synthesis in unseen data with a novel audio-text aligning compare to the expressive TTS with the autoregressive decoding.
    \item {\ours} achieves high expressivity and controllability through disentangled style factors with the same level of supervision compare to the non-autoregressive TTS.
    \item {\ours} is also noise-robust, proving that our novel noise modeling pipeline can decompose the noise as other style factors without any additional label.
\end{itemize}

%% file: 02_method_res.tex
\section{Method}

In this section, we describe our model first in an abstract level, then in detail. The noise modeling is introduced in the tail.

\subsection{Supervised Speech Decomposition}

There are two primary considerations to achieve our goal, non-autoregressive
TTS with style factor modeling. First, unlike the voice conversion task, the output speech's content is from text, and it may not be matched to the audio content. 
Second, recent non-autoregressive TTS frameworks benefit from supervision, while the speech decomposition is performed in unsupervised conditions. Solving these issues, we propose {\linearmelscaling} and a reinterpretation of speech decomposition under supervision.

\input{supplementary/model_architecture}
\subsubsection{\linearmelscaling}

Mismatch of length in text and audio has been solved by attention mechanism \cite{lee2019robust, sun2020fully}. Attention, however, has been claimed to be unstable, especially when synthesizing with unseen input data. In order to mitigate the limitation in robustness, we propose {\linearmelscaling}, a linear compression or expansion of the audio to the text's length. Only with the total lengths of both input text and audio, {\linearmelscaling} averages frames or repeats a frame of audio to assign them to a single phone by the ratio of the size of phoneme sequence over audio frame length. 
In this physically scaling method, the aligning process becomes a simple frame-wise bottleneck of incoming audio, requiring neither attention nor forced alignment. There are two advantages over the attention mechanism in terms of our objective: it does not bring any robustness issue against the non-autoregression, and it does not being exposed to the text so that audio-related style factors only come from the audio.  

\subsubsection{Information Bottleneck under Supervision}
\label{section:supervision}

Encoders of prior unsupervised speech decomposition \cite{qian2019autovc, qian2020speechsplit} need a tight bottleneck both channel-wise and frame-wise to deliver the necessary information from the source audio. But it makes training difficult and time-consuming. Under the supervision, encoders can utilize pre-obtained features from incoming audio such as pitch contour and energy. Since forced feature selection is not needed, excessive regulation will lead to information shortage and performance degradation \cite{qian2019autovc}. In our model, the bottleneck is adequately mitigated to fit the information flow relishing the supervision during style factor modeling.

\subsection{Model Architecture}

Figure~\ref{fig:figure1} shows an overview architecture of the proposed model. {\ours} models total six style factors: text, duration, pitch, speaker, energy, and noise. Text is regarded as a style factor equal to other audio related factors. To achieve rapidity, robustness, expressivity, and controllability at the same time, the elements of non-autoregressive TTS\footnote{https://github.com/ming024/FastSpeech2} \cite{ren2019fastspeech, ren2020fastspeech2} and speech decomposition approaches\footnote{https://github.com/auspicious3000/SpeechSplit} \cite{qian2019autovc, qian2020speechsplit} are adequately selected and transformed into {\ours}.

\subsubsection{Encoders}

Text encoder has two 256 dimensional Feed-Forward Transformer (FFT) blocks \cite{ren2019fastspeech} with four heads and takes phoneme sequence as input text. Since the text is encoded solely by the text encoder, the content of audio is removed through the bottleneck in other encoders. Duration, pitch, energy, and noise encoder have three 5x1 convolution layers followed by a group normalization \cite{wu2018groupnorm} of size 16, and channel-wise bottleneck by two layers bidirectional-LSTM of size 64 except the duration encoder where the size is 80. The dimension of the convolution layer is 256 for duration and noise, 320 for pitch and energy. {\linearmelscaling} is applied after the convolution stack. Encoder outputs are upsampled channel-wise before being sent to each predictor by linear layer with ReLU activation. Length regulator \cite{ren2019fastspeech, ren2020fastspeech2} is applied as a frame-wise upsampling to repeat frames for each phone counted by duration encoding.

The duration and noise encoder take mel-scaled spectrogram (mel-spectrogram). The pitch encoder takes a speaker normalized pitch contour (\(mean=0.5\), \(std=0.25\)) so that speaker-independent pitch contour is modeled. Speaker encoding is transferred from pre-trained speaker embedding, as presented in \cite{jia2018transfer}. In this work, Deep Speaker \cite{li2017deepspeaker} is employed since it is faithful to embed speaker identities without additional features such as noise robustness. The energy encoder takes scaled energy from 0 to 1 to ease the model training. After extracted from audio as in \cite{ren2020fastspeech2}, both pitch and energy input are quantized in 256 bins one-hot vector and then processed by encoders. Note that all inputs go through channel-wise bottleneck only. This is because the audio is already processed by the {\linearmelscaling}, and an additional frame-wise bottleneck acts as the excessive regulation discussed in Section~\ref{section:supervision}.

\subsubsection{Decoders}

{\ours} has three predictors \cite{ren2020fastspeech2}, each for the duration, pitch, and energy. The predictors need to consume all necessary information since they predict real values rather than latent codes. The input contains the sum of the corresponding encoding with text encoding. The text encoding is projected to four-dimensional space to balance the dependency. The pitch predictor receives speaker encoding as an additional input to predict the final pitch contour. We empirically find that adding speaker encoding to both downsampled and upsampled pitch encoding improves the decomposition of speaker identity.

Decoder, which predicts the final mel-spectrogram from disentangled style factors, has four 256 dimensional FFT blocks with four multi heads. The decoder's input is a combination of text encoding, pitch embedding (embedding of pitch predictor output), energy embedding (embedding of energy predictor output), and speaker encoding. Speaker encoding is included because information between input and target should be matched for the desired mapping.

\subsection{Noise modeling}
\label{section:noise}

On the way of improving the noise robustness of {\ours}, the noise can also be decomposed from audio without additional labels. In our model, noise is one of the remaining factors that have been explicitly encoded in model encoders, and it can be defined based on its residual property, excluding other style factors. In order to model noise by the definition, however, other encoders must be constrained to not include noise information even from the noisy input.

\subsubsection{Domain Adversarial Training}
\label{section:dat}
Not including noise information means extracting noise-independent features. In the TTS domain, this has been tackled by applying Domain Adversarial Training (DAT) \cite{ganin2016domain, hsu2019speakercorrelated}. Similar to previous works, the augmentation label and a gradient reversal layer (GRL) are introduced, which is jointly trainable in our model. The label of each predictor acts as a class label. As shown in Figure~\ref{fig:figure1}, DAT is applied to every audio-related encoding except noise encoder. After passing through GRL, each encoding is consumed by the augmentation classifier to predict the augmentation posterior (original/augmented). The augmentation classifier consists of two fully connected layers of 256 hidden size, followed by layer normalization \cite{ba2016layernorm} and ReLU activation. Note that each encoder except the noise encoder is now noise-independent, so each predictor's output is compared to clean labels rather than noisy ones.

\subsubsection{\doubledecoding}
\label{section:doubledecoding}
According to the definition of noise and the fact that an explicit label is not given, a novel pipeline called {\doubledecoding} is designed. It contains two phases: clean decoding and noisy decoding. In clean decoding, all noise-independent encodings are taken to predict a clean mel-spectrogrom, and in noisy decoding, noise encoder output is added to noise-independent encodings to predict a mel-spectrogrom with noise. Since only the noise encoder has to be updated at noisy decoding, the gradient does not flow through the other encoders. {\doubledecoding} can be seen as implicit supervision where the noise is directly compelled to focus on the leftover part of the audio with no explicit label.

\subsection{Loss Calculation}

The total loss of the model without noise modeling is: 
\begin{equation}
    Loss_{clean} = l_{mel-clean}~+~l_{duration}~+~l_{pitch}~+~l_{energy}
    \label{eq:cleanloss}
\end{equation}
where \(l_{mel-clean}\) is mean square error between predicted mel-spectrogram and target, and predictor losses (\(l_{duration}\), \(l_{pitch}\), and \(l_{energy}\)) are calculated by mean absolute error. Note that each extraction method of duration, pitch, and energy and the loss is the same as \cite{ren2020fastspeech2}, which means that the proposed model can disentangle style factors with no additional label.

The loss, including noise modeling, then becomes as follows: 
\begin{equation}
    Loss_{total} = Loss_{clean} ~+~ l_{mel-noisy} ~+~ l_{aug}
    \label{eq:noiseloss}
\end{equation}
where \(Loss_{clean}\) refers to Equation~\ref{eq:cleanloss}, \(l_{mel-noisy}\) is from noisy decoding calculated by the same way of \(l_{mel-clean}\) in Equation~\ref{eq:cleanloss}, and \(l_{aug}\) is the sum of each augmentation classifier loss in Section~\ref{section:dat}.

%% file: supplementary/model_architecture.tex
\begin{figure}[t]
  \centering
  \includegraphics[width=\linewidth]{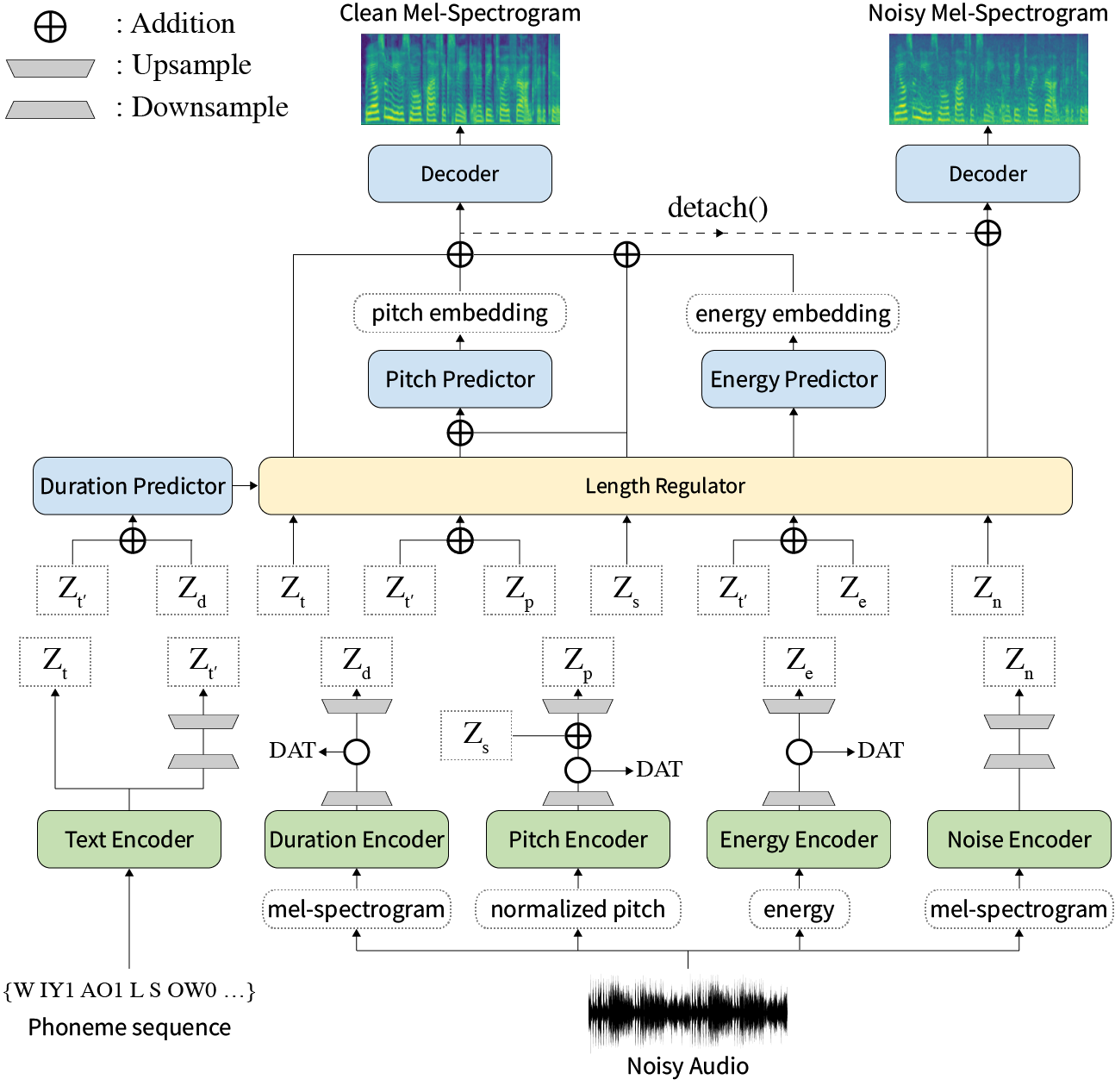}
  \caption{An architecture of the proposed model. \(\mathbf{Z}_\text{t}\), \(\mathbf{Z}_\text{t'}\), \(\mathbf{Z}_\text{d}\), \(\mathbf{Z}_\text{p}\), \(\mathbf{Z}_\text{s}\), \(\mathbf{Z}_\text{e}\) and \(\mathbf{Z}_\text{n}\) refer to text, downsampled text, duration, pitch, speaker, energy, and noise encoding, and DAT stands for Domain Adversarial Training.} 
  \label{fig:figure1}
\end{figure}

%% file: 03_experiment.tex
\section{Experiment}

\subsection{Experiment Setup}
Throughout the experiment, we train and evaluate models on VCTK corpus \cite{yamagishi2019cstr}. For the noisy dataset, we augment the original dataset by mixing each utterance with a randomly selected piece of background noise from WHAM! dataset \cite{Wichern2019WHAM}, at a random signal-to-noise ratio (SNR) in a range from 5 to 25. Raw audio is resampled at 22050Hz sampling rate and preprocessed to 1024 filter lengths, 1024 window sizes, and 256 hops lengths.

We divide the dataset into train, validation, and test set. In the first two sets, 44 speakers of each gender are randomly selected to balance speaker statistics. The remaining speakers are in the test set. The processed dataset is about 26 hours in total, containing 35805, 89, and 8345 utterances for the train, validation, and test set, respectively.

For the baseline, we use the following models to compare with {\ours} in four aspects of our goal.
\begin{itemize}
    \item Mellotron \cite{valle2020mellotron}: Tacotron2 GST \cite{shen2018natural, wang18gst} based autoregressive model, with ability to condition pitch and duration explicitly. We use this baseline to compare speed and robustness of expressive and controllable TTS.
    \item FastSpeech2 \cite{ren2020fastspeech2}: TTS with non-autoregressive decoding and supervision in pitch, duration, energy control. We select this baseline to compare expressiveness and controllability of reading style non-autoregressive TTS.
\end{itemize}

{\ours} is optimized by the same optimizer and scheduler in \cite{ren2020fastspeech2} with a batch size of 16. 
For the fair comparison, we apply pretrained ResCNN Softmax+Triplet model\footnote{https://github.com/philipperemy/deep-speaker} for the speaker embedding identically to all models. We use pretrained UNIVERSAL\_V1 model\footnote{https://github.com/jik876/hifi-gan} of HiFi-GAN \cite{kong2020hifi} as a vocoder.

\subsection{Evaluation Metrics}
We evaluate our model in four aspects: rapidity, robustness, expressiveness, and controllability. Training and inference time is strictly measured to prove the rapidity. Mean opinion score (MOS) is conducted to show the naturalness and robustness. We further investigate edge cases of weak robustness. Comparative MOS (CMOS) is conducted on a style transfer task to compare the expressiveness. For both MOS and CMOS, we use Amazon Mechanical Turk \cite{chu2006objective, ribeiro2011crowdmos}. Lastly, the controllability is demonstrated through an ablation study. There are two different speaker settings: seen speakers (S) and unseen speakers (US) in training time. And there are two different synthesis environments: parallel (P) indicates audio content and input text are identical, where nonparallel (NP) denotes the opposite.

\subsection{Results}

\subsubsection{Rapidity and Naturalness}
\input{supplementary/mos_speed_result}
Table~\ref{tab:result} shows experiment results of model speed and audio naturalness. For the model speed, we measure average training time per step during the first 1k steps and average inference time per single text from a set of speakers without vocoding. Even with increased times compared to FastSpeech2 due to the style factor modeling, {\ours} shows faster training and inference speed about \textbf{3.49\(\times\)} and \textbf{8.96\(\times\)} than Mellotron.

Comparing MOS score against FastSpeech2 confirms that {\ours} can synthesize natural speech even with increased model complexity. {\ours} is ranked higher than Mellotron for all conditions except S \& P where only a trivial gap exists. While both baselines show performance degradation under US or NP conditions, {\ours} shows better naturalness for both cases. In our auxiliary experiment, {\ours} begins to tilt on average style at a particular point of uncertainty, bringing more naturalness than expressive synthesis. It leads to higher MOS score in US than S with an increasing MOS score in US. In edge case investigation, Mellotron shows repeated and ignored words due to the broken alignment map, which is not the case in {\ours}. The naturalness tendency and edge cases indicate that our model synthesizes more natural speech than Mellotron especially when unseen data is consumed as input, proving unseen data robustness.

\subsubsection{Style Transfer}
\input{supplementary/cmos_result}
Table~\ref{tab:result2} reports CMOS of style transfer performance where {\ours} score is fixed to 0. Comparison with FastSpeech2 and Mellotron denotes that our method represents better expressiveness than both baselines. Also, the style transfer performance decreases in all conditions of Mellotron, showing our model's robustness on unseen data over the baseline. In another auxiliary experiment, there exists a trade-off between the naturalness and style transfer performance according to the dependency of each model input. Naturalness score, in detail, is proportional to the influence of text rather than audio, while style transfer performance is inversely proportional. FastSpeech2 is an extreme case where the model has text dependency only, and hence the best naturalness yet in average style. The phenomenon is relatively severe in duration so that the duration modeling could be less precise than other style factors. Note that even with such a trade-off, {\ours} outperforms Mellotron in all conditions.

\subsubsection{Style Factor Modeling}
Figure~\ref{fig:figure_res} shows our ablation study results as mel-spectrogram. Validating the functionality of style factor modeling, we synthesize speech while excluding each encoding one-by-one.

Synthesized from noisy reference audio (top-left), the resulting speech contains background noise (top-right) or not (middle-left) depending on whether the noise decoding is activated. Only the noise is synthesized when the noise encoding is solely activated (middle-right). From these results, it is proven that {\ours} can produce clean audio from the noisy reference audio, confirming our model's noise-robustness. We also find that noise modeling predicts the different noise levels of input audio. Summing up, these experiments conclude that our novel noise modeling successfully decomposes noise from reference audio like other style factors.

When the noise-independent style factors are ignored, the corresponding encodings become fixed style (bottom-left). Excluding pitch encoding, for instance, synthesizes speech has flat pitch contour. Also, given another speaker, the speaker's identity is changed while the other style factors are unchanged. When the text encoder is deactivated, audio-related style factors are modeled correctly, while synthesized speech is unintelligible (bottom-right). From all ablation study results, we conclude that encoders successfully capture expected style factors. 

In FastSpeech2, the duration, pitch, and energy are already controllable. However, the values are predicted from the same text and can be controlled only after being synthesized. In contrast, {\ours} distributes the dependency equally among text and style factors and controls them at the input levels. Therefore, with disentangled style factors, {\ours} can control output speech by input variants in addition to a single text, where all inputs can be of all different sources.

\input{supplementary/model_result}

%% file: supplementary/mos_speed_result.tex
\begin{table}[htb]
    \centering
    \caption{Time Measurement \& MOS Results}
    \begin{threeparttable}
        \resizebox{0.45\textwidth}{!}{
        \begin{tabular}{l|c|c|c|c|c|c}
            \toprule
            \multirow{2}{*}{Model} &
            \multicolumn{2}{c|}{Time (sec)} & \multicolumn{2}{c|}{S} & \multicolumn{2}{c}{US} 
            \\
            
             & Training & Inference & MOS-P & MOS-NP & MOS-P & MOS-NP
             \\
            \hline
            FastSpeech2 &
            0.102 & 0.019 & \multicolumn{2}{c|}{3.753} & \multicolumn{2}{c}{3.750} \\
            \hline
            Mellotron & 0.914 & 0.215 & 3.600 & 3.256 & 3.572 & 3.165 \\
            \hline
            \textbf{\ours} & \textbf{0.262} & \textbf{0.024} & \textbf{3.594} & \textbf{3.573} & \textbf{3.632} & \textbf{3.650} \\
            \bottomrule
        \end{tabular}
        }
    \end{threeparttable}
    \label{tab:result}
\end{table}

%% file: supplementary/cmos_result.tex
\begin{table}[h]
    \centering
    \caption{CMOS Results}
    \begin{threeparttable}
        \resizebox{0.4\textwidth}{!}{
        \begin{tabular}{l|c|c|c|c}
            \toprule
            \multirow{2}{*}{Model} & \multicolumn{2}{c|}{S} & \multicolumn{2}{c}{US} \\
             & CMOS-P & CMOS-NP & CMOS-P & CMOS-NP \\
            \hline
            \ours & 0 & 0 & 0 & 0 \\
            \hline
            FastSpeech2 & -0.303 & -0.478 & -0.388 & -0.266 \\
            \hline
            Mellotron & -0.334 & -0.422 & -0.25 & -0.359 \\
            \bottomrule
        \end{tabular}
        }
    \end{threeparttable}
    \label{tab:result2}
\end{table}

%% file: supplementary/model_result.tex
\begin{figure}[t]
  \centering
  \includegraphics[width=\linewidth]{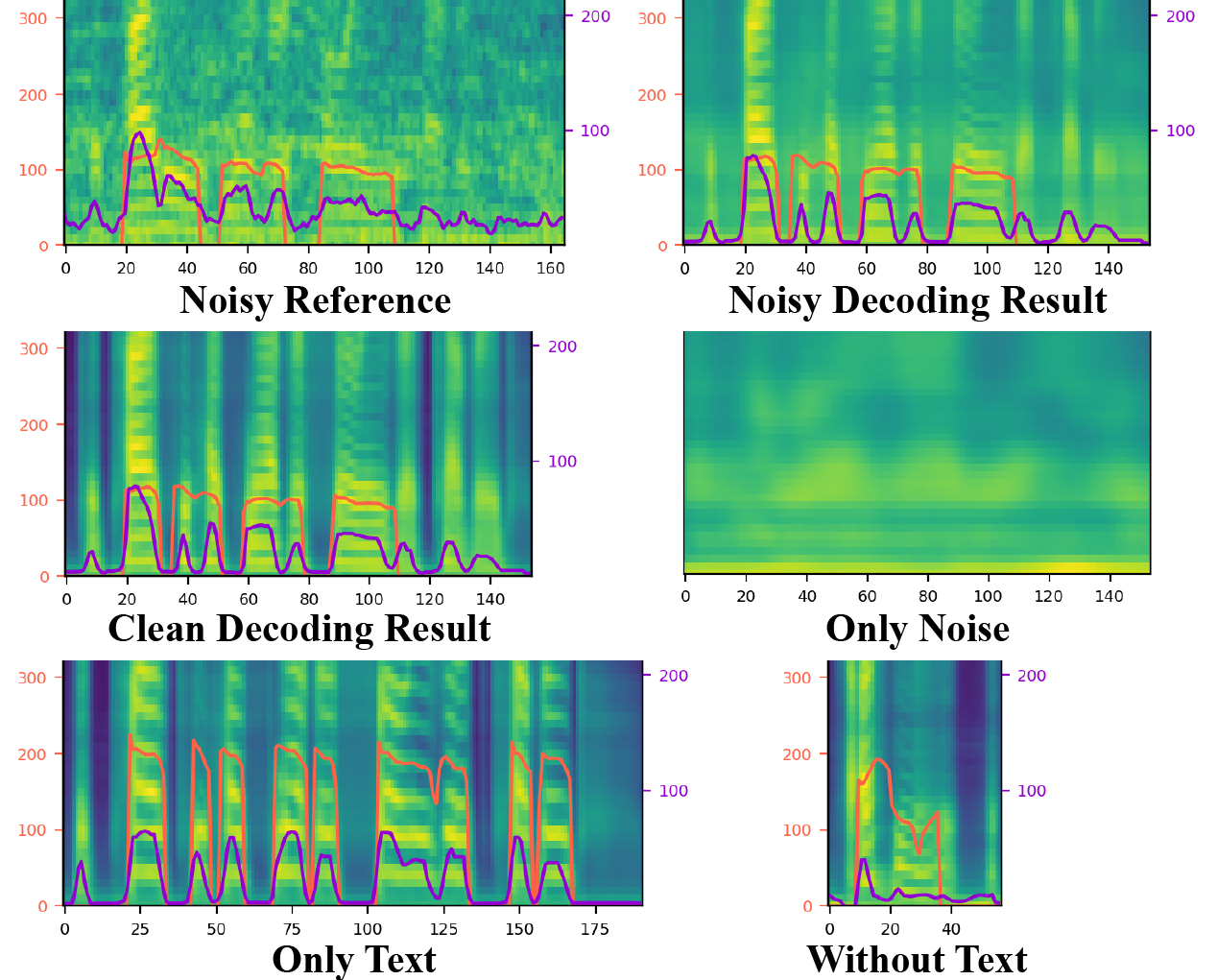}
  \caption{Results of style factor decomposition. The orange line represents pitch contour, and the purple line represents energy. The content is "The party has never fully recovered." and spoken by a male speaker. Both input text and audio are not shown in the training time. The figure is best shown in color.}
  \label{fig:figure_res}
\end{figure}


%% file: 04_conclusion.tex
\section{Conclusion}

In this paper, we propose STYLER, a non-autoregressive TTS framework with style factor modeling that achieves rapidity, robustness, expressivity, and controllability at the same time. It shows high performance and naturalness throughout various experiments. As discussed, however, our model has a trade-off between style transfer performance and speech naturalness. Our next plan is to resolve this trade-off, extending to different datasets and languages. We believe that {\ours} can become a cornerstone for applications and 
researches in TTS community. 

%% file: 05_ack.tex
\section{Acknowledgements}
This research was supported by the MSIT(Ministry of Science and ICT), Korea, under the Grand Information Technology Research Center support program(IITP-2021-2020-0-01489) supervised by the IITP(Institute for Information \& communications Technology Planning \& Evaluation) 